\documentclass[a4paper,12pt]{article}
\usepackage[T1]{fontenc}
\usepackage[letterpaper,tmargin=2.5cm,bmargin=2.5cm,rmargin=2.5cm,lmargin=2.5cm]{geometry}
\usepackage{amssymb,amsmath}
\usepackage{graphicx}
\usepackage{fancyhdr}
\usepackage{fancybox}
\usepackage{shadow}
\usepackage{empheq}
\usepackage{titlesec}
\usepackage{titletoc}
\usepackage{soul, color}
\usepackage{float}
\usepackage{shorttoc}
\usepackage{xcolor}
\usepackage{fancybox}
\usepackage{natbib}
\usepackage{array}
\usepackage{tabularx}
\usepackage{booktabs}
\usepackage{rotating}
\usepackage{bigstrut}
\usepackage{natbib}
\usepackage{aeguill}
\usepackage{tikz-qtree}
\everymath{\displaystyle}

\usepackage{changepage}

\usepackage{tikz}
\usetikzlibrary{calc}

\begin{document}
\title{Early Human Capital Accumulation and Decentralization}
\author{Guy Tchuente \thanks{%
School of Economics, University of Kent. Comments from Roland Pongou, Pierre NguimKeu and seminar participant from the Havard University are gratefully acknowledged.} \\
%EndAName
University of Kent \\
}
\date{ April 2021}
\maketitle
\abstract{Decentralization is a centerpiece in Cameroonian's government institutions' design. This chapter elaborates a simple hierarchy model for the analysis of the effects of power devolution. The model predicts overall positive effects of decentralization with larger effects when the local authority processes useful information on how to better allocate the resources.  The estimation of the effects of the 2010's power devolution to municipalities in Cameroon suggests a positive impact of decentralization on early human capital accumulation. The value added by decentralization is the same for Anglophone and Francophone municipalities; the effects of decentralization are larger for advanced levels of primary school.


}\newpage

\section{Introduction}

Decentralization and human capital are heavily investigated topics in the economics literature.\footnote{ \cite{channa2016decentralization} propose   a literature review on the importance of decentralization while, \cite{hanushek2008role} discusses the role of human capital for development.}  
 The geographical and ethnic diversity of Cameroon makes the invention of uniformly efficient institutions challenging.\footnote{ Cameroon has around 250 groups see \url{https://www.worldatlas.com/articles/ethnic-groups-of-cameroon.html} visited on 06/04/2021.}   The Cameroonian government has been implementing a regionalization process organized by Cameroon's Decentralization Law of 2019. The reliance on decentralization and democracy is based on studies establishing their crucial role for economic development (\cite{myersondemocratic} and \cite{myersonstrength}).  In Cameroon, causal empirical pieces of evidence on decentralization effects are rare.  This chapter proposed a methodological framework for the evaluation of decentralization policy and evaluated the effects of the 2010's power devolution in Cameroon on early human capital accumulation. 

In 2010, the Cameroonian government decentralized a substantial amount of input allocation decisions at the primary public school level from the central government to municipal authorities.  For example, the creation of schools, the rehabilitation of the classrooms, the construction of toilets, the water provision, and equipment maintenance are put under municipal authorities. In addition to the new competencies in infrastructure, each municipal authority is in charge of the allocation of the ``minimum package" to schools.\footnote{The ``minimum package" is an allocation of didactic material introduced in the year 2000 when primary public schools were declared free of charge by the government.} 

The assumption supporting this devolution of power by the central government is that proximity of the local authority will improve the allocation of the resources.  The interest in evaluating this decentralization policy is twofold. First, the 2010's decentralization policy is one of the first large-scale decentralization policies in Cameroon.  Second, it affects the supply side of the national instruments for early human capital production. Indeed, there are pieces of evidence in microeconomics and macroeconomics on the importance of the quality and quantity of human capital for the creation of wealth in low-income countries (\cite{hanushek2008role}, \cite{hanushek2012better}, and \cite{dahlum2017democracies}). Thus, evaluating the effect of the policy is crucial to understand the impact of future decentralization policy.

The economy of contract theory models the decentralization problems using hierarchical models. In these models, the principal (the central authority/ government) decides on whether  to delegate the decision power  to a supervisor (local authority/ municipality) on the contract to adopt with an agent (school).\footnote{\cite{mookherjee2006decentralization} provides a review of the use of mechanism design in an hierarchical model.}   This chapter uses a model with three types of  players: the government (the principal), the supervisor (municipality) and the agents (schools).  A school produces aggregated human capital as a function of basic effort exercised by pupils and resources allocated by the authority (government, municipality and schools).\footnote{The production function of human capital is similar to \cite{dal2020information} but the model differs with the role of the supervisor.}  The productivity of the resources allocated by the authority depends on how compatible they are with the needs of the school. Resources are assumed limited and  a uniform allocation is supposed when the government is in charge.  The municipality is allowed to have a more precise information on the needs of local schools and could, thus, realises improvement in the aggregated human capital production. The model predicts an overall positive effect of decentralization on human capital production. The effects are heterogeneous and increases with the ability of the local authority to better allocate resources. 

The predictions of the model are tested using data from Cameroon.   The government's power devolution affected directly pupils in public schools.  Thus, public school pupils are considered as the treatment group in this quasi-experiment while the private school's pupils are the control group. The year 2005 observations are gathered before policy intervention while the 2014 observations are collected after the policy intervention. A difference-in-differences methodology, based on PASEC\footnote{Programme d'Analyse des Systemes Educatifs de la Confemen.} 2005 and 2014 pupil's mathematics, and literacy ability test scores data, is used to evaluate the policy intervention. The ability of the difference-in-differences design to identify a policy-relevant parameter is explained and the effects of the 2010's decentralization policy are estimated using the ordinary least squared (OLS) with different specifications.

The estimation of the effects of the decentralization policy suggests a positive effect of decentralization on early human capital accumulation.  The devolution of power to municipalities has increased the tests score of primary school pupils by 10.20\% in mathematics and 15.39\% in literacy.  The decomposition of the effect by level shows that the effect is driven by changes at the end of the cycle level. Indeed,  the estimation suggests small policy effects for Grade 2 pupils. This is a confirmation of the prediction of the theoretical model. Indeed, it will be difficult to clearly know the specific need of pupils at such an early level, while the overall school needs can be better assessed by the municipality.  Thus, around the end of the cycle (Grades 5 or 6), the pupils have enough time to benefit from better-targeted resources allocation by the municipal authority. The decomposition along the linguistic line suggests a small difference in the effect of the decentralization between the Anglophone and the Francophone primary school system. This may be driven by the fact that there is virtually no difference between the information that can be acquired by Francophone and Anglophone municipalities. 

Human capital is accounted as a main contributor to wealth creation for countries (see \cite{manuelli2014human} and \cite{schoellman2016early} ). This chapter contributes to the literature assessing how government institutions affect human capital production by analyzing how decentralization in a low-income country. It, thus, adds to contributions such as \cite{davies2003educational} who study decentralization in Malawi's education sector, while \cite{pomuti2012decentralization} investigate the role plays by decentralization policy in Namibia.  There are some empirical pieces of evidence suggesting that decentralized supervision improves the professional development of teachers (see \cite{esia2014effects} for the study of Ghana,  \cite{pomuti2012decentralization} in Namibia, and \cite{tamukong2004towards} presenting a decentralization model for Cameroon). But, some studies find that decentralization may have negative effects on funding allocation, with the largest portion of funding not reaching the target population (see \cite{otieno2013effectiveness} and \cite{dembele2015financement}).  In a review of the literature on decentralization in education and health proposed by \cite{channa2016decentralization} none of the studies of decentralization in the educational sector is considered to have a ``Very strongly credible identification strategy".    Moreover, among the studies deemed as having a ``Strongly credible identification" none of them are from an African country. This chapter aims to implement a rigorous quasi-experimental design to evaluate the effect of decentralization in Cameroon.   Thus, for Cameroonian researchers and policymakers, this chapter develops a simple framework that can guide them in the process of implementation and evaluation of future decentralization policies. 

The remaining chapter is organized as follows. Section 2 presents some background information on the primary education sector in Cameroon as well as some useful knowledge regarding 2010's power devolution. Section 3 presents a simple model of decentralization in the context of human capital production.   Section 4 discusses the empirical strategy used to evaluate the effect of the decentralization policy and tests the implications of the model. It also explores the estimation results and presents relevant policy implications. A brief conclusion follows in Section 5.

\section{Background and Data}

The development of the education system in Cameroon since 1990 has been oriented by international commitments and local ambitions. At the international level, the Education For All (EFA) conference of Jomtien, Thailand (1990), organized by UNESCO and the Dakar's (2000) meeting where several countries expressed their commitment to the Millennium Development Goals (MDG) are the most prominent.  At the local level, the policies are implemented to fit government growth and poverty reduction strategies as express in the Poverty Reduction Strategy Paper (PRSP, 2003)and the Growth and Employment Strategy Paper (GESP, 2009).  The government vision for education consists of providing the youths with quality education, requisite competencies, and professional attitudes that will facilitate their insertion and competitiveness in the professional world (\cite{alemnge_curriculum_2019}). The education system is divided into four stages:  pre-primary (4 to 5 years old), primary (6 to 11 years old), secondary (12 to 18 years old), Tertiary (19 to 23 years old). This chapter evaluates the effect of decentralization on primary public schools in Cameroon. 

\subsection{Cameroon Primary Schools: A Brief Overview}

The primary school service in Cameroon has three main providers: the government (public schools), private, and private confessional (Catholic schools, Islamic schools, protestant schools, etc). Each official language (French and English) has its independent primary school system. Both systems are represented in all regions of the country with a predominance of French schools in the French-speaking region, while English schools form the majority of English-speaking regions (North-West and South-West regions) schools. According to the \cite{pasec2014}'s report on the state of education in Cameroon, during the academic year 2014-2015, the francophone system has 71.6\% of pupils while the anglophone has 28.4\%. Among pupils, the share attending public institutions is around  75\% overall. This share is larger in the francophone system (79\%) compare to the Anglophone system (62\%). 

%%Comments on the trend Primary school indicators
\begin{figure}[!h] 
\centering
\includegraphics[width=.7\linewidth]{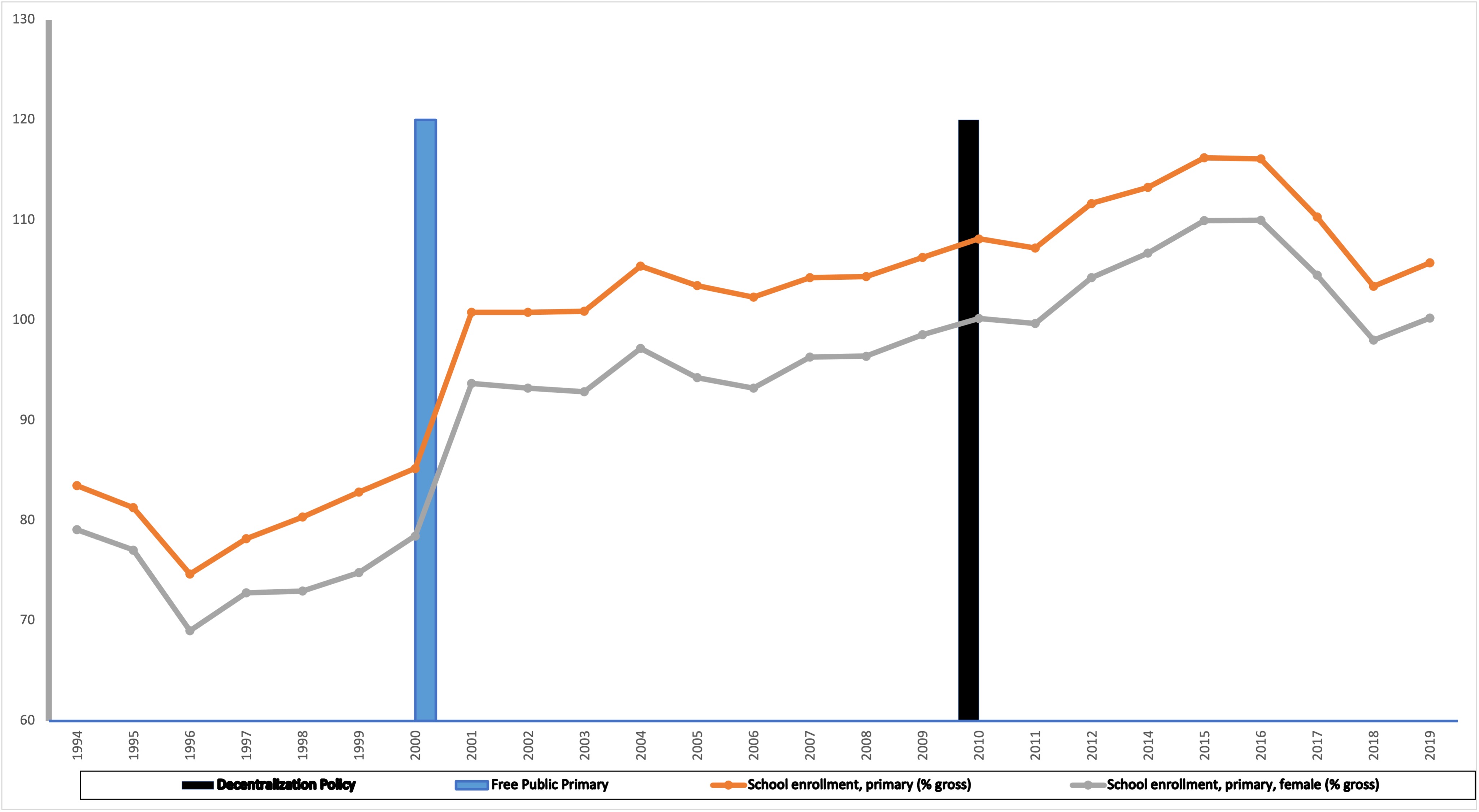}
\caption{Enrolment Rate 1994 to 2019 from the World Bank data.}\label{Enroltrend}
\end{figure}

Figure \ref{Enroltrend} shows a steady increase in the primary school enrolment rate between 1996 and 2016 for boys and girls. Accelerations of the enrolment rate are notable in 2000 and 2010.  These years correspond respectively to the year of the announcement of the free public primary school by the government and the devolution of substantial allocation powers to the municipal level.  
%%Comments on curriculum policy
Since the independence, the Francophone and the Anglophone systems have studied with different curricula with some harmonization policy by the government.  These efforts have led to a joint curriculum effective for implementation from the 2018/2019 school year. The intention to harmonize the systems has been manifested by the government since the days of independence and materialized by the 1960 law on the organization of education. After these early steps, in 1990  the government launched its first national curriculum the ``New pedagogical approach". After nearly a decade of implementation of the ``new pedagogical approach", the government started, between 1998 and 2000, a new syllabus based on the National Forum on Education held 1995.  Following that reform, in 2003 the government decided to implement a  pedagogic approach called the Competence-Based Approach (CBA), in all primary schools of the country \cite{alemnge_curriculum_2019}. 
%%No big changes in the supply side since 2005

Between 2005 and 2018 there have not been any major change in the primary school curriculum in Cameroon.\footnote{It is important to note that all schooling the country must follow the official curriculum.} The supply of education services has been marked during the same period by steady recruitment of teachers in the Anglophone and Francophone systems (the ratio of pupils per teacher was 47.8 in 2005 and 45.8 in 2014). The most important change in the period between 2005 and 2014 is a decentralization policy applied by the government to all public schools.  

\subsection{The 2010's Power Devolution Policy }

The public primary education in Cameroon was inherited from colonial times with a very high level of centralization. Sine 1990, the Cameroonian government has adopted decentralization as an instrument to achieve the Education For All goal.  The commitment to apply decentralization in education was reinforced by several laws.  The year 2010 marks a substantial step in the implementation of the government decentralization ambitions. The municipalities are granted the right to allocate resources to public primary schools in their area of authority.  For example, the municipality has to right to create schools (following the government's school map guidance). They are in charge of the maintenance of public primary school equipment, they recruit support staff for the school. They also participate in the acquisition of didactic material. This includes the provision to the public schools of the ``minimum package" necessary for the functioning of the schools.\footnote{See \url{http://bibliotheque.pssfp.net/index.php/institutions/institutions-nationales/senat/les-collectivites-territoriales-decentralisees/les-decrets/951-decret-n-2010-0247-pm-du-26-fevrier-2010-fixant-les-modalites-dexercice-de-certaines-competences-tranaferees-par-letat-aux-communes-en-matirer-education-de-base/file}, visited on 08/04/2021, for the details of the decentralization decision.} 

Alongside the resource allocation power, the municipalities were granted the right to produce public school infrastructure: the construction and rehabilitation of classrooms, construction of latrine blocks, water points and fences, maintenance and equipment of public nursery and primary schools are the responsibility of the municipality. Figure \ref{Enroltrend} suggests that the 2010 devolution of power was followed by an acceleration in the enrolment rate in primary schools.

\section{A Simple Model of Hierarchy}

The decentralization problem can be represented using a hierarchical model where the government is the principal, the municipality is the supervisor, and the public schools are the agents (there are $N$ schools in the municipality). 

Each school is a unit mass agent producing human capital $h_i$. The production of human capital is a function of $$h_i(l_i)=l_is_i+e_i$$ where $l_i$ is the resource allocated by the municipality, $s_i$ measures the compatibility of the resource allocated to the school $i$ with its needs and, $e_i$ is the level of effort made by the school. All the variables have bounded support.  

In a private school, the human capital production is only a function of the school effort ($e_i$) as they receive do not rely on the government or municipality to finance their enterprise.  For public schools, the resource granted by the municipality or the government could a positive or a negative effect on human capital production. This means that $s_i$ can be either negative or positive. 

 Let $l_{i0}$ be the initial resources endowment of school $i$ and $l_{i1}$ the resource endowment after a small increase. Let $\Delta l_i= l_{i1}-l_{i0}>0$, the human capital production is
 \begin{equation} \label{prodfun}
   h_i= e_i + s_il_{i0}+ s_i \Delta l_i  
 \end{equation}

In general, $\Delta l_i$ additional resource will produce $s_i \Delta l_i$ additional units of human capital. 

As the compatibility of resources with local needs is not expected under central government allocation. Let assume a uniform increased in allocation to schools.  $s_i, e_i$ are considered independent and identically distributed across all schools.  The expected increase in human capital $$\delta= E(s_i \Delta l_i).$$ Under central government,  $\Delta=\Delta l_i$. Thus,  $\delta= E(s_i)\Delta$. The total additional resource to be allocated is given $N\Delta$ where $N$ is the number of schools.  The quantity $N\Delta$ is the total amount of additional resources available that can be granted to the municipality for allocation. 

The local authority (municipality) can choose the distribution of extra resources such that $s_i\Delta l_i\geq s_i\sum_{i=1}^N \Delta l_i /N$ and $\sum_{i=1}^N \Delta l_i=N\Delta$.

Let consider the set of allocation $M_i=\{ s_i\Delta l_i\geq s_i\sum_{i=1}^N \Delta l_i /N and \sum_{i=1}^N \Delta l_i=N\Delta \}$.  These allocations will improve production of individual with the larger compatibility factor ($s_i$). $M_i$ is a no-empty set in presence of decentralization. It can be shown that

\begin{equation}\label{central_rel}
    \delta= E(s_i)\Delta  \leq E(s_i \Delta l_i| M_i)=\rho_i.
\end{equation}

The quantity $\lambda_i=(\rho_i-\lambda)$ measures the expected gain from decentralization. The inequality in Equation (\ref{central_rel}) implies the following points:
\begin{enumerate}
    \item An additional unit of resource allocated under decentralization is at least as productive as under centralization.
    
    \item If the central authority is able to meet the local schools' needs the decentralization will bring no profit. 
    
    \item The effect of decentralization is heterogeneous and depends on the ability of the authority to distinguish different types of schools ($s_i$).   
\end{enumerate}
 
The model predicts an overall positive effect of decentralization on human capital production. To test the predictions of our model, an ideal solution would have been to run a randomized control trial (RCT).  This will imply the creation of representative treatment and control groups of municipalities. The schools in the treatment group are put under decentralization and those in the control leave in the status quo. This chapter takes an alternative route. It takes advantage of the quasi-experimental nature of the power devolution to municipalities in 2010 to measure the causal effect of decentralization on human capital accumulation using a difference-in-difference design.

\section{Empirical Investigation of the 2010's Decentralization}
\subsection{Research Design: Difference-in-differences}

In 2010 the Cameroonian government decided to implement a decentralization policy. This policy affected public primary schools in several aspects. However, private and private confessional schools were not affected by the policy. Let consider public primary schools as the treatment group ($P_i=1$) and private schools as the control group ($P_i=0$).\footnote{Given the difference in the production function of human capital, they may not be suitable control as pupils may select in different school depending on their unobserved characteristics.}

The human capital outcome of pupils from both groups is observed before and after 2010. Consider $dT=1$ if an observation is collected in 2014 and $dT=0$ otherwise.  This allows the estimation of the average treatment effect on the treated (ATT) of decentralization.  

Let $h_{it}$ be the human capital variable of a school/pupil $i$ and time $t$. The model predicts that

\begin{equation} \label{Cont_fac}
   h_{it}= e_{it} +(\rho_i-\delta) dT P_i  + s_{it}  l_{i0t} +E(s_{it})  \Delta - E(s_{it} \Delta l_{it}) + s_{it} \Delta l_{it}
\end{equation}
 where $dT$ is a dummy indicating if the observation is in a period with decentralisation. 

Let consider $\lambda_i=(\rho_i-\delta)$ be the benefit from decentralization.  Based on the model, if $P_i=0$  then $M_{it}$ is empty and $\rho_i$ is not defined. However, it is possible to define $\lambda_0=E(\lambda_i| P_i=1)$. This is the ATT of decentralization. 

Let $X_{it}=(Z_{it}, W_{it})$ be the set of exogenous characteristics. $Z_{it}$ is a set of individual characteristics affecting the school efforts $e_{it}=e(Z_{it},\varepsilon_{0it})$.  $W_{it}$  are schools features related to the compatibility of resources $s_{it}=s(W_{it},\varepsilon_{1it})$. The assumption of linearity of $s$ and $e$ implies that
\begin{equation} \label{Cont_fac_diff}
   h_{it}= Z_{it}'\theta_1 +(\rho_i-\delta) dT P_i  +W_{it}'\theta_2  l_{i1t} +\theta_0  + \varepsilon_{0it} + \varepsilon_{1it}l_{1it}
\end{equation}

Let assume that $l_{1it}=\mu_0 +\mu_1 dT +\mu_2 P_i +\varepsilon_{2it}$. The structural Equation (\ref{Cont_fac_diff}) is approximated by
\begin{equation} \label{Cont_fac_diff1}
   h_{it}= \tilde{\mu}_1 dT +\tilde{\mu}_2 P_i+ Z_{it}'\theta_1 +(\rho_i-\delta) dT P_i  +W_{it}'\tilde{\theta}_2  +\theta_0  + \varepsilon_{0it} + \varepsilon_{1it}(\mu_0 +\mu_1 dT +\mu_2 P_i +\varepsilon_{2it}).
\end{equation}
The transformed structural equation is 

\begin{equation} \label{Cont_fac}
   h_{it}= \beta_0+X_{it}'\beta_1 + \alpha_0 P_i + \alpha_1 dT +  \lambda_0 dT \times P_i +v_{it}.
\end{equation}

with $v_{it}= \varepsilon_{it}+(\lambda_i-\lambda_0)dT \times P_i$ where $\varepsilon_{it}=\varepsilon_{0it} + \varepsilon_{1it}(\mu_0 +\mu_1 dT +\mu_2 P_i +\varepsilon_{2it})$.  An OLS estimator of $\lambda_0$ is unbiased if $v_{it}$ is uncorrelated with $dT\times P_i$ under the assumption the extra resources allocation is not selective based on unobserved school characteristic ($E(\varepsilon_{1it}\varepsilon_{2it}|P_i ,dT, X_{it})=0$).  %This implies the parallel trend assumption. 

The set-up described above corresponds to a difference-in-differences design. The following section presents the data and discusses the parallel trend assumption in the PASEC sample. 

\subsection{Data and Identification Assumption Verification}

The data used for the empirical investigation are 2005 and 2014 school survey data from the Programme d’Analyse des Systèmes Educatifs de la CONFEMEN (PASEC) for Cameroon.
The is a representative sample of pupils from schools in French and English- speaking Cameroon.
The 2005 sample has pupils from Grade 2 and  Grade 5 classes (15 pupils are then randomly drawn, interviewed, selected, and tested in a class of each grade). In 2014 another random sample is drawn from   Grade 2 and  Grade 6. The tests administrated by the PASEC are in language and maths. They are designed to allow comparison over time and across systems. Alongside the test score,  data contains school characteristics, socioeconomic characteristics of the pupils, their parents, and some useful household information.

A crucial assumption of our empirical strategy is that in absence of decentralization in 2010, the private and public schools' pupil's scores in language and maths would have had the same trend. This assumption is not directly testable. The curriculum followed by private schools is the same as public schools. Also, there has not been another major policy intervention in the sector of public primary schools. This coupled with the heterogeneity of private school practice, implies that there is a very high chance of validation of the parallel trend assumption. %However, this assumption is tested by verifying the difference between private and private confessional schools.\footnote{The results for their regression show no effect for maths but a negative effect on literacy.} 

\subsection{Empirical Results and Policy Implications }

The estimation of the decentralization effects is carried out using the model represented in Equation (\ref{Cont_fac}).   The accumulation of early human capital is measured by mathematics and literacy test scores.  The model's parameters are estimated with and without school fixed-effects and on the sub-sample of Grade 2 test scores. 

The main parameter of interest is  $\lambda_0$.  It measures the effects of decentralization of early human capital accumulation in Cameroon. Our results suggest a positive effect of decentralization.  The models that explain most of the variability in the human capital accumulation in the full sample are models with school fixed-effects (FE).  Thus, these are the preferred specifications. They are in column Math with FE and Lit. with FE.  These results suggest that decentralization has increased Math scores by 10.2\% and Literacy scores by 15.39\%.  These effects are significant both statistically and economically. 

The columns Math Gr 2, and Lit. Gr2,  are the estimation of the parameters using the sub-population of Grade 2 pupils. The estimation of the decentralization effects suggests small but statistically significant effects.  The relatively low-level pupils may make a better allocation of recourse to this particular group more challenging to the municipality. This interpretation is in line with the finding of our model of Section 3 suggesting no difference between the central and decentralized regimes when the municipality cannot target the best match for the resources. 

Expect for literacy in Grade 2, decentralization has a positive effect on all the human capital accumulation measures.  Thus, the allocation of the municipality could be view to be at least as good as the central authority allocation in our sample. 

As Cameroon has the particularity of being divided into two school systems, namely the anglophone and francophone system, Table 2 presents the assessment of the heterogeneity of the decentralization effect along that dimension. The result suggests small but statistically significant difference between francophone and anglophone.  Decentralization seems to have a relatively small effect effect in Anglophone Municipalities compare to Francophone ones. However, the difference in score is below 5\%. It also worth-nothing that, everything being equal, Anglophone pupils have roughly 10\% higher scores than a pupil from the Francophone system. Thus, the small relative better performance of Francophone municipality could come from the fact that they had more room fro growth.  
Also, there is no reason to expect the municipality to better match the needs of pupils in the anglophone system compare to the francophone one, thus, the model in Section 3 predicts similar decentralization effects. 

The positive effects of decentralization suggested by the regression results demonstrate the importance of the proximity of the decision-making to the population.  These results should motivate the government to accelerate the decentralization process prescribed by the Cameroon's Decentralization Law of 2019.  This chapters' model and empirical findings suggest that decentralization should prioritized sectors where the local authorities can have better quality information on the production of the goods and services.

\begin{table}[htbp]\centering
\def\sym#1{\ifmmode^{#1}\else\(^{#1}\)\fi}
\caption{Estimation of the Decentralization Effect}
\resizebox{14cm}{10cm}{\begin{tabular}{l*{6}{c}}
\hline\hline
                    &\multicolumn{1}{c}{Math }&\multicolumn{1}{c}{Lit. }&\multicolumn{1}{c}{Math }&\multicolumn{1}{c}{Lit.}&\multicolumn{1}{c}{Math Gr2}&\multicolumn{1}{c}{Lit. Gr2}\\
\hline

After Treatment &      -2.759\sym{***}&      -3.424\sym{***}&      -16.44\sym{***}&      -14.80\sym{***}&      -22.23\sym{***}&      -19.30\sym{***}\\
                    &     (-3.71)         &     (-4.19)         &    (-53.68)         &    (-46.67)         &    (-77.35)         &    (-74.60)         \\
[1em]
Public Schools       &      -6.617\sym{***}&      -6.160\sym{***}&      -11.93\sym{***}&      -16.35\sym{***}&       1.146         &      -7.590\sym{***}\\
                    &     (-8.61)         &     (-7.42)         &    (-28.41)         &    (-35.23)         &      (1.27)         &     (-9.51)         \\
[1em]
$\lambda_0$   &       1.705\sym{*}  &       0.825         &       10.20\sym{***}&       15.39\sym{***}&      -4.848\sym{***}&       5.780\sym{***}\\
                    &      (2.12)         &      (0.94)         &     (19.71)         &     (27.43)         &     (-5.45)         &      (7.57)         \\
[1em]
Anglophone          &      -1.094\sym{***}&      -3.261\sym{***}&       9.388\sym{***}&       10.88\sym{***}&       7.110\sym{***}&       10.22\sym{***}\\
                    &     (-3.42)         &     (-9.75)         &     (23.54)         &     (24.40)         &      (9.29)         &     (14.38)         \\
[1em]
Age                 &      -0.551\sym{***}&      -1.090\sym{***}&     -0.0295         &      -0.273         &       0.560\sym{*}  &       0.129         \\
                    &     (-5.27)         &    (-10.03)         &     (-0.22)         &     (-1.91)         &      (2.11)         &      (0.50)         \\
[1em]
Dummy for Girl      &      -0.539         &       0.466         &      -0.680\sym{*}  &       0.118         &      -0.548         &       0.657         \\
                    &     (-1.78)         &      (1.48)         &     (-1.99)         &      (0.37)         &     (-0.78)         &      (1.12)         \\
[1em]
Books at Home       &       3.374\sym{***}&       3.068\sym{***}&       1.394\sym{**} &       1.971\sym{***}&      -0.217         &       0.552         \\
                    &      (9.97)         &      (8.79)         &      (2.87)         &      (4.19)         &     (-0.25)         &      (0.77)         \\
[1em]
Electricity at Home  &       2.198\sym{***}&       2.488\sym{***}&      -0.949\sym{*}  &      -1.362\sym{**} &       1.723\sym{*}  &       1.166         \\
                    &      (6.43)         &      (6.76)         &     (-2.33)         &     (-2.88)         &      (2.00)         &      (1.48)         \\
[1em]

Dummy for higher grades          &      -7.493\sym{***}&      -4.901\sym{***}&      -10.16\sym{***}&      -9.422\sym{***}&                     &                     \\
                    &    (-13.16)         &     (-8.34)         &     (-9.14)         &     (-8.44)         &                     &                     \\
                    \hline
School Fixed-effects        &    No        &       No         &       Yes        &       Yes        &       Yes         &       Yes         \\
\hline
Observations        &       10078         &       10078         &       10078         &       10078         &        4076         &        4076         \\
\(R^{2}\)           &       0.127         &       0.131         &       0.393         &       0.433         &       0.434         &       0.551         \\
\hline \hline
\multicolumn{5}{p{13.5cm}}{\scriptsize{\textbf{Notes:} statistics in parentheses.  \sym{*} \(p<0.05\), \sym{**} \(p<0.01\), \sym{***} \(p<0.001\). The outcome variables are pupils' test scores in mathematics and literacy in 2005 and 2014. The year 2014 is the year after the treatment and public schools pupils are in the treatment group, pupils are either in Grade 2, 5 or 6. Standard errors are clustered at the school level.}}
\end{tabular}}
\end{table}

\begin{table}[htbp]\centering
\def\sym#1{\ifmmode^{#1}\else\(^{#1}\)\fi}
\caption{Estimation of the Decentralization Effect Anglophone vs. Francophone}
\resizebox{11cm}{9.5cm}{\begin{tabular}{l*{4}{c}}
\hline\hline
                    &\multicolumn{1}{c}{Math }&\multicolumn{1}{c}{Lit. }&\multicolumn{1}{c}{Math Gr2}&\multicolumn{1}{c}{Lit. Gr2}\\
\hline
 After Treatment&      -16.44\sym{***}&      -14.80\sym{***}&      -22.23\sym{***}&      -19.30\sym{***}\\
                    &    (-53.68)         &    (-46.67)         &    (-77.35)         &    (-74.60)         \\
[1em]
Public School       &      -11.93\sym{***}&      -16.35\sym{***}&       1.146         &      -7.590\sym{***}\\
                    &    (-28.41)         &    (-35.23)         &      (1.27)         &     (-9.51)         \\
[1em]
$\lambda_0$   &       10.20\sym{***}&       15.39\sym{***}&      -4.848\sym{***}&       5.780\sym{***}\\
                    &     (19.71)         &     (27.43)         &     (-5.45)         &      (7.57)         \\
[1em]
Decent $\times$ Anglophone &      -4.163\sym{***}&      -4.819\sym{***}&       2.352\sym{***}&      -1.964\sym{***}\\
                    &     (-9.93)         &    (-11.01)         &     (20.13)         &    (-19.76)         \\
[1em]
Anglophone          &       9.388\sym{***}&       10.88\sym{***}&       7.110\sym{***}&       10.22\sym{***}\\
                    &     (23.54)         &     (24.40)         &      (9.29)         &     (14.38)         \\
[1em]
Age                 &     -0.0295         &      -0.273         &       0.560\sym{*}  &       0.129         \\
                    &     (-0.22)         &     (-1.91)         &      (2.11)         &      (0.50)         \\
[1em]
Dummy for Girl      &      -0.680\sym{*}  &       0.118         &      -0.548         &       0.657         \\
                    &     (-1.99)         &      (0.37)         &     (-0.78)         &      (1.12)         \\
[1em]
Books at Home       &       1.394\sym{**} &       1.971\sym{***}&      -0.217         &       0.552         \\
                    &      (2.87)         &      (4.19)         &     (-0.25)         &      (0.77)         \\
[1em]
Electricity at Home  &      -0.949\sym{*}  &      -1.362\sym{**} &       1.723\sym{*}  &       1.166         \\
                    &     (-2.33)         &     (-2.88)         &      (2.00)         &      (1.48)         \\
[1em]
Dummy for higher grades            &      -10.16\sym{***}&      -9.422\sym{***}&                     &                     \\
                    &     (-9.14)         &     (-8.44)         &                     &                     \\

\hline
School fixed effects        &       Yes        &       Yes        &       Yes        &       Yes         \\
\hline
Observations        &       10078         &       10078         &        4076         &        4076         \\
\(R^{2}\)           &       0.393         &       0.433         &       0.434         &       0.551         \\
\hline\hline
\multicolumn{5}{p{12.5cm}}{\scriptsize{\textbf{Notes:} statistics in parentheses.  \sym{*} \(p<0.05\), \sym{**} \(p<0.01\), \sym{***} \(p<0.001\). The outcome variables are pupils' test scores in mathematics and literacy in 2005 and 2014. The year 2014 is the year after the treatment and public schools pupils are in the treatment group, pupils are either in Grade 2, 5 or 6. Standard errors are clustered at the school level.} }
\end{tabular}}
\end{table}

\section{Conclusion and Related Future Issues}

This chapter has investigated the impact of the 2010 decentralization in the Cameroonian education sector on early human capital accumulation.  This policy is one of the major government recent actions in the primary education section. 

A simple model of hierarchy is proposed to represent the government decentralization problem. The theoretical model suggests that when the municipality is able to achieve a better allocation of resources, decentralization has a positive impact. It also suggests heterogeneity in the returns to decentralization. 

Data from PASEC containing pupils' test scores in math and literacy, measures to ensure comparability, are used to estimate and test the predictions of the model. The application of a difference-in-differences design enables the estimation of the causal effect of decentralization.  Indeed, decentralization has a positive effect on early human capital accumulation. The effect of power devolution in 2010 is larger for literacy (15.39\%) than in mathematics (10.2\%).  These effects are only noticeable at the end of the primary cycle. The also no difference between the decentralization effect for Anglophone and Francophone systems.

\newpage
\bibliography{product_network}
\bibliographystyle{econometrica}

\end{document}